\documentclass[pre,aps,showpacs,nofootinbib,twocolumn]{revtex4}
\usepackage{amssymb}
\usepackage{graphicx}

\input{tcilatex}

\begin{document}

\title{ ARPES Spectra of Bi2212 give the Coulomb
Coupling $\lambda ^{C}\approx 1$\\ and the Electron-Phonon
Coupling $\lambda ^{EP}=2-3$}
\author{Miodrag L. Kuli\'{c}$^{1}$, Oleg V. Dolgov$^{2}$}

\address{$^{1}$Max Planck Institute for Physics of Complex Systems, 01187 Dresden,
Germany\\
$^{2}$Max Planck Institute for Solid State Physics, 70569
Stuttgart, Germany}

\begin{abstract}
We show that the double kink-structure in the electronic self-energy of
Bi2212 near the nodal point at low energy $\omega _{1}\approx 50-70$ $meV$
and at high energy at $\omega _{2}\approx 350$ $meV$, observed recently in
the ARPES measurements by Valla et al \cite{Valla-2006}, gives that the
electron-phonon (EPI) coupling constant $\lambda _{z}^{EP}$in the normal
part of the self-energy $\Sigma (\omega )$ is twice larger than the Coulomb
coupling $\lambda _{z}^{C}$. The experimental data for $\func{Re}\Sigma
(\omega )$ up to energies $\sim 350$ $meV$ can be satisfactory explained by $%
\lambda _{z}^{EP}\approx 2.1$ and $\lambda _{z}^{C}\approx 1.1$.
Additionally the low energy slope of the ARPES $\func{Re}\Sigma (\omega )$
at $\omega <20$ $meV$ \cite{Valla-2006} gives a hint that the low energy
phonons might contribute significantly to the EPI coupling, i.e. $\lambda
_{z}^{low,EP}>1$, thus giving the total EPI coupling constant $\lambda
_{z,tot}^{EP}=\lambda _{z}^{EP}+\lambda _{z}^{low,EP}>3$. In order to test
the role of low frequency phonons by ARPES measurements a much better
momentum resolution is needed than that reported in \cite{Valla-2006}.
Possible pairing scenarios based on ARPES, tunnelling and magnetic neutron
scattering measurements are discussed.
\end{abstract}

\date{\today }
\maketitle

In the last several years ARPES measurements in high temperature
superconductors (HTSC), with much better momentum and energy resolution,
made a breakthrough in determining the energy and momentum dependence of the
quasiparticle self-energy in HTSC. Important measurements by the Shen group
\cite{Shen-review}, \cite{Shen-Cuk-review} gave evidence on the low-energy
kink in the quasiparticle spectrum around the phonon energy between 40-70
meV, in \emph{both nodal and antinodal points}. Since the low-energy kink is
pronounced practically in all HTSC materials, above and below T$_{c}$, it is
most probably of the phononic origin, as it was recently confirmed by the
ARPES isotope effect in $\func{Re}\Sigma $ \cite{Lanzara-isotope}. These
results gave a new impetus in the HTSC physics by renewing the importance of
the electron-phonon interaction (EPI) in the quasiparticle scattering and in
the pairing mechanism of HTSC. In spite of the very convincing evidence in
ARPES, tunnelling, STM and optic measurements that EPI in HTSC is rather
strong \cite{Maksimov-review}, \cite{Kulic-review}, \cite%
{Kulic-Dolgov-review}, \cite{Kulic-Ginzburg issue}, the proponents of the
spin-fluctuation ($SF$) mechanism of pairing still persist to explain main
features in the electronic self-energy \emph{solely} by this non-phononic
coupling.

One of the central questions in the HTSC physics is - which of these two
mechanisms, EPI and SF, contributes dominantly in the low energy region -
relevant for pairing. To remind the reader, the strength of the
quasiparticle scattering in the normal part of the self-energy ($\Sigma (%
\mathbf{k},\omega )$) in the Migdal-Eliashberg theory (approximation) can be
characterized by the total coupling constant
\begin{equation}
\lambda _{z}(\mathbf{k})=2\dsum\limits_{i}\int_{0}^{\infty }\frac{\alpha _{%
\mathbf{k},i}^{2}F_{i}(\omega )}{\omega }d\omega ,  \label{lambda}
\end{equation}%
where the summation goes over all bosonic modes involved in the scattering
mechanism. Note, that at low energies usually one has $\Sigma (\mathbf{k}%
,\omega )=-\lambda _{z}(\mathbf{k})\omega $ for $\omega \ll \omega _{b}$,
where $\omega _{b}$ is the (smallest) characteristic bosonic frequency. We
stress the well known fact, that the Migdal-Eliashberg theory is well
defined for the electron-phonon (boson) scattering mechanism with $\lambda
\Omega /W\ll 1$, where $\Omega $ is the characteristic phonon (boson) energy
and $W$ is the band width. However, in the case of the SF mechanism there is
\textit{no controllable Migdal-Eliashberg theory}. Therefore, the proponents
of the $SF$ mechanism assume a \emph{phenomenological form} for the spectral
function $\alpha _{\mathbf{k}}^{2}F(\omega )_{SF}\sim g_{SF}^{2}\func{Im}%
\chi (\mathbf{k},\omega )$, where $\chi (\mathbf{k},\omega )$ is the spin
susceptibility. In systems where $\func{Im}\chi (\mathbf{k},\omega )$ is
strongly peaked around some $\mathbf{k}$-points, like in underdoped HTSC
where $\mathbf{k}=\mathbf{Q}_{AF}(\pi /a,\pi /a)$, this approximation for $%
\alpha _{\mathbf{k}}^{2}F(\omega )_{SF}$ is unwarranted, since the vertex
correction (terms beyond the Migdal-Eliashberg approximation) can
significantly influence the self-energy which is most probably suppressed
\cite{Schrieffer}, \cite{Hanke}.

It is worth of mentioning, that in order to fit the quasiparticle
self-energy by the SF theory it is necessary to make a radical assumption
that the SF coupling constant is (unrealistically) large, i.e. $g_{SF}\sim 1$
$eV$ \cite{MMP}. Such a large value of $g_{SF}$ is difficult to justify,
both theoretically and experimentally. In Ref. \cite{Levin} it was shown
that the SF phenomenology fails to give large $T_{c}$ even for $g_{SF}\gg 1$
$eV$, if the spectral function $\func{Im}\chi (\mathbf{k},\omega )$ is taken
from the neutron scattering data and not from the low frequency NMR spectra
as it was done in \cite{MMP}. The situation is even worse if one fits the
slope $d\rho /dT$ of the resistivity $\rho (T),$ which gives $g_{SF}\sim 0.3$
$eV$ \ and $T_{c}<7$ $K$ - see discussions in \cite{Maksimov-review}, \cite%
{Kulic-review}, \cite{Kulic-Dolgov-review} and references therein. The most
impressive evidence for ineffectiveness of SF to explain (solely) the
self-energy effects came from the magnetic neutron scattering measurements
by the Bourges's group \cite{Bourges-neutron}. They show, that by changing
doping from slightly underdoped to optimally doped systems there is a
dramatic change in the magnitude and $\omega $-shape of the magnetic
spectral function $\func{Im}\chi (\mathbf{k\approx Q},\omega )$ in the
normal state at $T>T_{c}$. These experiments show \cite{Bourges-neutron}
that AT $T>T_{c}$ in the \emph{slightly underdoped} $YBa_{2}Cu_{3}O_{6.92}$
(with $T_{c}=91$ $K$) $\func{Im}\chi (\mathbf{Q},\omega )$ is peaked (with
\emph{very large value})\emph{\ }at $\omega \approx 35$ $meV$, while in the
\emph{optimally doped}  $YBa_{2}Cu_{3}O_{6.97}$ (with $T_{c}=92.5$ $K$) $%
\func{Im}\chi (\mathbf{Q},\omega )$ is \emph{drastically suppressed} and
practically negligible. This result was confirmed quite recently in Ref.
\cite{Reznik} in the optimally doped systems where $\func{Im}\chi (\mathbf{k}%
,\omega )$ is practically negligible at all $\mathbf{k}$. Moreover, in spite
of the large suppression of $\func{Im}\chi (\mathbf{k},\omega )$ in the
optimally doped HTSC \cite{Bourges-neutron} there is a \emph{very small
difference in} $T_{c}$ ($91$ $K$ vs $92.5$ $K$) \emph{of slightly underdoped
and optimally doped HTSC}. This means that the large reconstruction and
suppression of $\func{Im}\chi (\mathbf{k},\omega )$ by doping \emph{does not
affect superconductivity at all}. This result means also that $g_{SF}\ll 1$ $%
eV$ in the SF phenomenology and that the latter should be abandoned as the
leading pairing mechanism in HTSC. It is also worth of mentioning that the
recent numerical calculations on Hubbard and t-J model render valuable
evidence that there is \emph{no high temperature superconductivity} (SC) in
these models \cite{Kivel}. If SC exists at all in these models its $T_{c}$
must be rather low. Therefore, other interactions, such as EPI and the
direct Coulomb interaction, should be taken into account.

\begin{figure}[tbp]
\begin{center}
\resizebox{.6 \textwidth}{!} {
\includegraphics*[width=7cm]{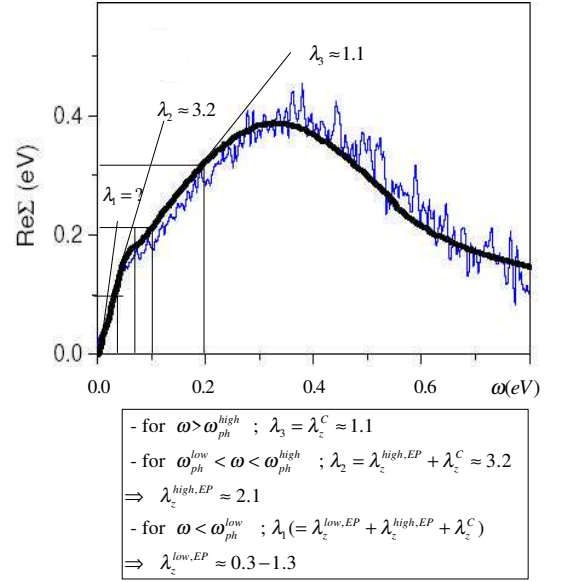}}
\end{center}
\caption{ Fig. 4b from \protect\cite{Valla-2006}: $\func{Re}\Sigma (\protect%
\omega )$ measured in Bi2212 (thin line) and model $\func{Re}\Sigma (\protect%
\omega )$ (bold line) obtained in \protect\cite{Valla-2006}. The three thin
lines ($\protect\lambda _{1},\protect\lambda _{2},\protect\lambda _{3}$) are
the slopes of $\func{Re}\Sigma (\protect\omega )$ in different energy
regions.}
\label{Scen-fig}
\end{figure}

In Refs. \cite{Kul-Dolg-ARPES}, \cite{MDK} it was argued that the low-energy
(below $1$ $eV$) behavior of the ARPES $\Sigma (\mathbf{k},\omega )$,
obtained recently in \cite{Shen-review}, \cite{Shen-Cuk-review}, \cite%
{Lanzara-isotope}, can be qualitatively and semi-quantitatively explained by
the \textit{combined effect }of the EPI and Coulomb interaction. The recent
ARPES results \cite{Shen-review}, \cite{Shen-Cuk-review}, \cite%
{Lanzara-isotope} for the effective real part of the self-energy at energies
$\omega <\omega _{ph}^{\max }\approx 80$ $meV$ ($\omega _{ph}^{\max }$ is
the maximal phonon frequency) give evidence that the \emph{effective EPI
coupling }\cite{Shen-Cuk-review} is rather strong $\sim 1$\textit{. }%
However, in \cite{Shen-review}, \cite{Shen-Cuk-review}, \cite%
{Lanzara-isotope} the EPI self-energy was obtained by subtracting the high
energy slope of the quasiparticle spectrum $\omega (\xi _{k})$ at $\omega
\sim 0.3$ $eV$. The latter is due to the Coulomb interaction. Although the
position of the low-energy kink is not affected by this procedure (if $%
\omega _{ph}^{\max }\ll \omega _{C}$), the above (subtraction) procedure
gives in fact an \textit{effective EPI self-energy }$\Sigma _{eff}^{EP}(%
\mathbf{k},\omega )$ and\textit{\ coupling constant} $\lambda _{z,eff}^{EP}(%
\mathbf{k})$ only. Let us briefly demonstrate that $\lambda _{z,eff}^{EP}(%
\mathbf{k})$ is smaller than the real EPI coupling constant $\lambda
_{z}^{EP}(\mathbf{k})$. The total self-energy is $\Sigma (\mathbf{k},\omega
)=\Sigma ^{EP}(\mathbf{k},\omega )+\Sigma ^{C}(\mathbf{k},\omega )$ where $%
\Sigma ^{C}$ is the contribution due to the Coulomb interaction. At very low
energies $\omega \ll \omega _{C}$ one has usually $\Sigma ^{C}(\mathbf{k}%
,\omega )=-\lambda _{z}^{C}(\mathbf{k})\omega $, where $\omega _{C}(\sim 1$ $%
eV)$ is the characteristic Coulomb energies and $\lambda _{z}^{C}$ the
Coulomb coupling constant. The quasiparticle spectrum $\omega (\mathbf{k})$
is determined from the condition%
\begin{equation}
\omega -\xi (\mathbf{k})-\func{Re}[\Sigma ^{EP}(\mathbf{k},\omega )+\Sigma
^{C}(\mathbf{k},\omega )]=0,  \label{Disp}
\end{equation}%
where $\xi (\mathbf{k})$ is the bare band structure energy. At low energies $%
\omega <\omega _{ph}^{\max }\ll \omega _{C}$ Eq.(\ref{Disp}) can be
rewritten in the form%
\begin{equation}
\omega -\xi ^{ren}(\mathbf{k})-\func{Re}\Sigma _{eff}^{EP}(\mathbf{k},\omega
)=0,  \label{Low-Disp}
\end{equation}%
where $\xi ^{ren}(\mathbf{k})=[1+\lambda _{z}^{c}(\mathbf{k})]^{-1}\xi (%
\mathbf{k})$ and $\func{Re}\Sigma _{eff}^{EP}(\mathbf{k},\omega )=[1+\lambda
_{z}^{c}(\mathbf{k})]^{-1}\func{Re}\Sigma ^{EP}(\mathbf{k},\omega )$. Since
at very low energies $\omega \ll \omega _{ph}^{\max }$one has $\func{Re}%
\Sigma ^{EP}(\mathbf{k},\omega )=-\lambda _{z}^{EP}(\mathbf{k})\omega $ and $%
\func{Re}\Sigma _{eff}^{EP}(\mathbf{k},\omega )=-\lambda _{z,eff}^{EP}(%
\mathbf{k})\omega $, then the real coupling constant is related to the
effective one by
\[
\lambda _{z}^{EP}(\mathbf{k})=[1+\lambda _{z}^{C}(\mathbf{k})]\lambda
_{z,eff}^{EP}(\mathbf{k})>\lambda _{z,eff}^{EP}(\mathbf{k}).
\]%
At higher energies $\omega _{ph}^{\max }<\omega <\omega _{C}$, which are
less important for pairing and where the EPI effects are suppressed, one has
$\func{Re}\Sigma (\mathbf{k},\omega )\approx -\lambda _{z}^{C}(\mathbf{k}%
)\omega $. The conclusion is that in order to obtain the correct values for
the EPI and Coulomb coupling constant the self-energy $\func{Re}\Sigma (%
\mathbf{k},\omega )$ should be measured in a broad energy interval. This was
done in the recent ARPES measurements on Bi2212 and LSCO by Valla et al.
\cite{Valla-2006} where $\func{Re}\Sigma (\mathbf{k},\omega )$ was
experimentally determined in the broad energy interval. The measured $\func{%
Re}\Sigma ^{\exp }(\mathbf{k},\omega )$ at $T=10$ $K$ near and slightly away
from the \emph{nodal point} in the optimally doped Bi2212 with $T_{c}=91$ $K$
\cite{Valla-2006} is shown in Fig. \ref{Scen-fig}.

It is seen in Fig. \ref{Scen-fig} that $\func{Re}\Sigma ^{\exp }(\mathbf{k}%
,\omega )$ has \textit{two kinks} - the first one at \textit{low energy} $%
\omega _{1}\approx \omega _{ph}^{high}\approx 50-70$ $meV$ which is most
probably of the phononic origin \cite{Shen-review}, \cite{Shen-Cuk-review},
\cite{Lanzara-isotope}, while the second kink at \textit{higher energy} $%
\omega _{2}\approx \omega _{C}\approx 350$ $meV$ is evidently due to the
Coulomb interaction (by including spin fluctuations too). However, one of
the most important results in Ref. \cite{Valla-2006} is that the slopes of $%
\func{Re}\Sigma ^{\exp }(\mathbf{k},\omega )$ at low ($\omega <\omega
_{ph}^{high}$) and high energies ($\omega _{ph}^{high}<\omega <\omega _{c}$)
are \textit{very different}. The low-energy and high-energy slope \emph{near
the nodal point} are depicted and shown in Fig. \ref{Scen-fig} schematically
(thin lines). From Fig. \ref{Scen-fig} it is obvious that EPI prevails at
low energies $\omega <\omega _{ph}^{high}$. More precisely digitalization of
$\func{Re}\Sigma ^{\exp }(\mathbf{k},\omega )$ in the interval $\omega
_{ph}^{high}<\omega <0.4$ $eV$ gives the Coulomb coupling $\lambda _{z}^{C}$
\begin{equation}
(\lambda _{z}^{SF}<)\lambda _{z}^{C}\approx 1.1  \label{lamda-c}
\end{equation}%
while the same procedure at $20$ $meV\approx \omega _{ph}^{low}<\omega
<\omega _{ph}^{high}\approx 50-70meV$ gives the total coupling constant $%
(\lambda _{2}\equiv )\lambda _{z}=\lambda _{z}^{EP}+\lambda _{z}^{C}\approx
3.2$ and the EPI coupling constant $\lambda _{z}^{EP}(\equiv \lambda
_{z}^{high,EP})$

\begin{equation}
\lambda _{z}^{EP}\approx 2.1\text{.}  \label{lambda-exp}
\end{equation}%
This estimation tells us that at and near the nodal point \textit{the EPI
interaction dominates }in the quasiparticle scattering at low energies since
$\lambda _{z}^{EP}(\approx 2.1)\approx 2\lambda _{z}^{C}>2\lambda _{z}^{SF}$%
, while at large energies (compared to $\omega _{C}$) the Coulomb
interaction with $\lambda _{z}^{C}\approx 1.1$ dominates. We point out that
EPI near the anti-nodal point can be even larger than in the nodal point,
mostly due to the higher density of states at the anti-nodal point.

It is well known that the low energy quasiparticle scattering dominates the
transport and thermodynamic properties of metallic systems such as HTSC.
Comparing the Valla et al. results \cite{Valla-2006} with the previous ARPES
measurements \cite{Shen-review}, \cite{Shen-Cuk-review} it is apparent that
the real EPI coupling constant $\lambda _{z}^{EP}(\mathbf{k})$ - obtained
from experiments in Ref. \cite{Valla-2006}, are at least twice larger than
the effective one $\lambda _{z,eff}^{EP}(\mathbf{k})$ - from Refs. \cite%
{Shen-review}, \cite{Shen-Cuk-review}, \cite{Lanzara-isotope}, i.e. $\lambda
_{z}^{EP}(\mathbf{k})\approx 2\lambda _{z,eff}^{EP}(\mathbf{k})$.

Moreover, the results for $\func{Re}\Sigma ^{\exp }(\mathbf{k},\omega
)(=-\lambda _{1}\omega )$ at \textit{very low energies} $\omega <\omega
_{ph}^{low}\approx 20$ $meV$, shown in Fig. \ref{Scen-fig}, give a \textit{%
hint} to an even larger slope which gives the low-energy coupling $\lambda
_{1}=\lambda _{z}^{low,EP}+\lambda _{z}^{EP}+\lambda _{z}^{C}$. This slope
is larger than those in the (higher energy) interval $20$ $meV<\omega <80$ $%
meV$ \cite{Valla-2006} which means that the EPI coupling due to low energy
phonons may be rather strong, i. e. $\lambda _{z}^{low,EP}\gtrsim 0.3-1.3$,
while the total EPI coupling is $\lambda _{z,tot}^{EP}=\lambda
_{z}^{low,EP}+\lambda _{z}^{EP}>2.4-3.4$ - we call this the \textit{%
L-scenario}.

It is worth of pointing out that if the \textit{L-scenario} turns out be
correct and if the high value of $\lambda _{z}^{low,EP}\gtrsim 1.3$ is
realized, then the \emph{vibrations of heavier atoms} (than oxygen) may
contribute significantly to pairing in $HTSC$. One of the consequence of
this result would be a reduction (from the canonical value $0.5$) of the
\emph{oxygen isotope effect} in optimally doped $HTSC$ materials, as it was
observed experimentally\ - see \cite{Kulic-review} and references therein.
Furthermore, the value $\lambda _{z}^{low,EP}\gtrsim 1.3$ is compatible with
the earlier tunnelling measurements on $Bi_{2}Sr_{2}CaCu_{2}O_{8}$ (with $%
T_{c}\approx 70$ $K$) \cite{Shimada} which give the total EPI coupling
constant $\lambda _{tunn}^{EP}\approx 3.5$, while $\lambda
_{tunn}^{low,EP}\approx 2.1$ for $\omega <20$ $meV$, as well as with Ref.
\cite{Zhao} where the tunnelling experiments from Ref. \cite{Wei} were
analyzed. However, to elucidate the role of low energy phonons by ARPES
measurements a much better momentum resolution ($\delta k$) is needed than
that reported in \cite{Valla-2006}, where $\delta k\approx 0.04$ \AA $^{-1}$%
\ and the self-energy resolution is $\delta \func{Re}\Sigma >20$ $meV$. This
resolution is insufficient for a definitive conclusion on the contribution
of low energy phonons.

Finally, although it is still premature for giving a definitive pairing
scenario in HTSC, the experimental evidence for d-wave pairing and for the
strong EPI in HTSC imply a necessary condition for EPI in HTSC:\textit{\ EPI
in HTSC must be peaked at small transfer momenta}, i.e. there is \textit{%
forward scattering peak }in EPI. Otherwise, if the rather large EPI would be
weakly momentum independent it would inevitably destroy d-wave pairing in
HTSC - see more in \cite{Kulic-review}, \cite{Kulic-Dolgov-review}, \cite%
{Kulic-Ginzburg issue}. In that respect several pairing scenarios are
imaginable depending on the strength of the EPI coupling in the \emph{d-wave
channel} $\lambda _{d,tot}^{EP}(<\lambda _{z,tot}^{EP})$. Since the SF
coupling is small and $\lambda _{z}^{SF}\ll \lambda _{z}^{C}\sim 1$, then if
$\lambda _{d,tot}^{EP}$ is sufficiently strong ($\lambda _{d,tot}^{EP}>1$)
than d-wave pairing is dominated by EPI which in conjunction with the
Coulomb interaction (by including both the short- and long-range parts) my
give high $T_{c}$. In that case the s-wave part $\lambda _{s}^{C}$ of the
Coulomb interaction suppresses s-wave pairing while the d-wave part $\lambda
_{d}^{C}(\ll \lambda _{s}^{C})$ if $\lambda _{d}^{C}>0$ affects d-wave
pairing weakly. If the Coulomb interaction is attractive in the d-channel
\cite{Leggett}, i.e. $\lambda _{d}^{C}(<0)$, it strengthens d-wave pairing
additionally. Very interesting situation arises if both EPI and the (total)
Coulomb interaction give appreciable attraction in the d-channel with $%
\lambda _{d,tot}^{EP}\sim \mid \lambda _{d}^{C}\mid $. The reasons for the
large EPI coupling constant in the d-wave channel have been discussed in
\cite{Kulic-review}, \cite{Kulic-Dolgov-review}, where it was argued that
strong correlations in conjunction with the weakly screened EPI Madelung
coupling in the ionic-metallic structure of HTSC can produce the forward
scattering peak in the EPI coupling and other charge scattering processes
\cite{Kulic-Zeyher}, \cite{Kulic-review}. In such a case the EPI coupling in
the d-wave channel $\lambda _{d}^{EP}$ may be appreciable, while on the
other hand the transport coupling due to EPI ($\rho (T)\sim \lambda _{tr}T$)
is suppressed, i.e. $\lambda _{tr}^{EP}<\lambda _{z}^{EP}/3$ \cite%
{Kulic-Zeyher}, \cite{Kulic-review}. The latter result resolves the
long-standing experimental puzzle in \ HTSC that the experimental value of $%
\lambda _{tr}$ is too small to give high $T_{c}$. Contrary to low
temperature superconductors, where in most materials $\lambda _{tr}\approx
\lambda _{z}\approx \lambda _{z}^{EP}$, in HTSC materials one has $\lambda
_{tr}\ll \lambda _{z}$ due to the forward scattering peak in EPI \cite%
{Kulic-Zeyher}, \cite{Kulic-review}. If the spin fluctuation spectral
function would give $\lambda _{d}^{SF}\approx \lambda _{z}^{C}$ then $SF$
and $EPI$ would interfere constructively giving high $T_{c}$. However, as we
discussed above this scenario contradicts the magnetic neutron scattering
measurements of the Bourges's group \cite{Bourges-neutron}, \cite{Reznik},
which imply that $\lambda _{d}^{SF}\ll \lambda _{z}^{C}$. Therefore, the SF
approach should be abandoned.

In conclusion, we have argued that the recent ARPES measurements of the
self-energy in Bi2212 at and near the nodal point by Valla et al. \cite%
{Valla-2006} give evidence that at energies $\omega <70$ $meV$ the EPI
interaction is at least twice stronger than the Coulomb interaction (which
includes spin fluctuations too). It turns out that $\lambda _{z}^{EP}>2.1$
and $\lambda ^{SF}<\lambda _{z}^{C}\approx 1.1$. These ARPES measurements
give also a hint that the low-energy phonons can give an appreciable EPI
coupling $\lambda ^{low,EP}\gtrsim 1$ and $\lambda _{z,tot}^{EP}>3$. In
order to clear up the role of low frequency phonons in EPI ARPES
measurements need much better resolution in momentum space than that
reported in \cite{Valla-2006}. If confirmed by other experiments the ARPES
results by Valla et al. \cite{Valla-2006} will render undisputable evidence
for the leading role of the electron-phonon interaction in the scattering
and pairing mechanism of HTSC materials.

\textbf{Acknowledgement} - We thank Adrian Gozar for useful discussions on
the ARPES data. M. L. K. thanks Peter Fulde for support.

\end{document}